\documentstyle[epsfig,12pt]{article}
\begin{document}
\begin{center}
{\Large Analysis of hadron production
in nucleus-nucleus interactions
up to and out of kinematical
limit of free NN-collisions in the frame of
FRITIOF model}
\vskip 0.5cm
{\Large A.S. Galoyan, G.L. Melkumov, V.V. Uzhinskii }\\
  JINR,
 Dubna, Russia
\date{}
\end{center}

\begin{center}
 \begin{minipage}{5.2in}
   {\it In the framework of the modified FRITIOF model, the inclusive
     spectra of the cumulative $\pi ^0$-, $\pi ^- $-mesons and protons
     produced  in the nucleus-nucleus
     interactions at 4.5 GeV/c/nucleon  and 4.2 GeV/c/nucleon are
     calculated. It is shown  that the model reproduces qualitatively,
     and in some cases quantitatively  the main experimental
     regularities of $\pi $-mesons production, and "soft" part of
     the proton spectra.  According to the model the production of the
     cumulative particles is connected with the mechanism of the "soft"
     nucleon-nucleon interaction.}
\end{minipage}
\end{center}

The production  of particles in  hadron-nucleus and nucleus-nucleus
interactions in the kinematical region forbidden for free hN-collisions
(so-called cumulative processes) is one of the main sources of
information about the nuclear structure function and the multi-particle
production mechanism. This kinematical region corresponds to the
Feyman variable $X_F > 1 $ (with respect to a single
$NN$-interaction).
$  X_F=2 p_l^*/\sqrt{S}$,
where $p_l^*$
is the longitudinal momentum,   $\sqrt{S} $ is the total energy
of NN-interaction in the nucleon-nucleon center of mass system.

The cumulative phenomena predicted by A.M. Baldin \cite{Baldin-Pred}
were discovered by the group of Stavinski \cite{Stavin-eff} in early
seventies, and have been intensively investigated during last thirty
years. The wide experimental and theoretical study caused numerous
models for explanation of this interesting phenomenon.
The main property of all models is the existence of massive (heavier
than nucleon) compact object on which the creation of cumulative
particle is occurred. In dependence on this object creation
all models can be divided into the "hot" \cite{Goren} and "cold"
 \cite{Strikman}, \cite{Lukyaniv} models.
The models for explanation of the cumulative processes have analytical
form  and can  pretend on a description of restricted kinematical
region.The reproduction of their results are   quite
complicated.

At the same time, there are created the phenomenogical Monte Carlo
generators of events with  easily reproduced results. But the
cumulative production of particles isn't assumed directly in the
 code-generators.
  It is needed to note that "hot" models ideas have found an
extraordinary application in high energy physics in some modified
forms. They are used in the well-known models of multi-particle
production such as FRITIOF \cite{FRITIOF}, RQMD \cite{RQMD}, and HIJING
\cite{HIJING}. The common assumption of the models is that the soft
inelastic hadron-hadron collisions have a binary character $ a~ + b~
\rightarrow a'~ + b'~$, where $a'$  and $b'$ are excited hadrons. The
excited hadrons with masses $m_{a'}$, $m_{b'} >~ m_a$, $m_b$ are
considered as QCD-strings, and LUND-model \cite{JETSET} is used to
describe their decays.

In the case of hadron-nucleus interactions, the models assume that an
excited hadron $a'$ can collide with other nuclear nucleons and
increase its mass. The same can take place in nucleus-nucleus
interactions.

As one can easily notice, that the general representation of the
hadron-nucleus interactions  assumed by the models is almost similar to
that considered in Ref. \cite{Goren}. The authors of Ref. \cite{Goren}
supposed that heavy hadron system (a fireball) which does not include
a leading particle is created in the first collision of projectile
hadron with a nuclear nucleon. The fireball moving in the nucleus
collides with other nucleons, slows down, and increases its mass. As a
results, a production of particles in the regions kinematically
forbidden in free hadron-nucleon collisions becomes possible. Thus, one
can expect that the cumulative particles have to appear in the models,
in particular, in the FRITIOF model.

Figures 1, 2 show the experimental data \cite{EXP}
on  fast $\pi ^0$- mesons production in nucleus-nucleus interactions at
$P=$ 4.5 A GeV/c with FRITIOF model calculations taking into account
the last corrections \cite{Gana}. As seen, the FRITIOF model predicts
the cumulative particle production.
\begin{figure}[cbth]
\begin{center}
\psfig{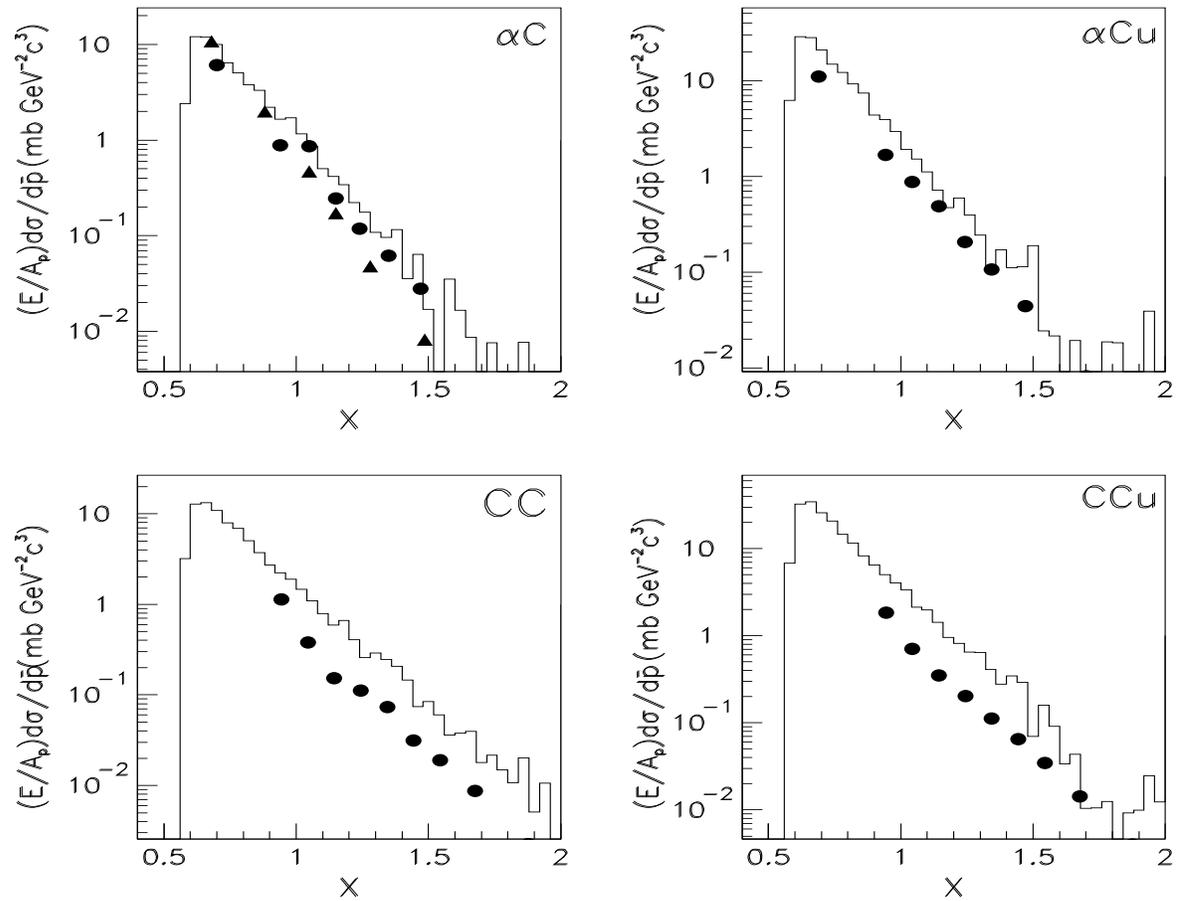}
\caption{Invariant inclusive cross-sections of $\pi ^0$-meson production
in nucleus-nucleus interactions at 4.5 GeV/c/nucleon. The points are experimental data
of the FOTON setup, histograms are the FRITIOF model calculations.}
\end{center}
\label{fig1}
\end{figure}

 $\pi ^0$- mesons production in the
 $pC$-, $pCu$-, $\alpha C$-, $\alpha Cu$-, $CC$- and $C
Cu$- interactions at momentum   4.5 $A_p$ GeV/c have been studied
experimentally in Refs. \cite{EXP}.  $\gamma$- quanta were registered
at the experiments by 90-channel Cherenkov $\gamma$-spectrometer
of LHE FOTON setup.   $\pi ^0$- mesons with  the angles
in laboratory  system (in the rest frame of  target )
$\theta _{\pi} \leq 16^0$ and energies  $E_{\pi} \geq 2$ GeV
were considered after estimation background conditions and
$\pi ^0$ identification.

In fig. 1, the experimentally measured invariant
 cross-sections
of  $\pi ^0$- mesons production  per mass number of projectiles ($A_p$)
as a function of cumulative number  $X$ are presented by circles.
The variable $X$ was determined  as
$$
X=\frac{m_N E_{\pi ^0} - m^2_{\pi^0}/2}{E_N m_N - E_N E_{\pi^0} - m^2_N
+ P_N P_{\pi ^0} cos \theta _{\pi ^0}},
$$
where $m_N$ and $m_{\pi ^0}$ are nucleon and meson masses,
respectively, $P_N$ is the momentum of projectile per nucleon  ($P_N =
4.5$ GeV/á).
$P_{\pi ^0}$ is  $\pi ^0$ momentum,  $E_N=\sqrt{M_N^2 + P_N^2}$,
$E_{\pi ^0} = \sqrt{m_{\pi ^0}^2 + P_{\pi ^0}^2}$.
The systematic errors of  the cross-section is about $\sim $ 20 \%.
The statistical errors are in the circle limits.

In fig. 1, histograms demonstrate calculations of the
cross-sections of  $\pi ^0$- mesons production at $E_{\pi ^0} \geq 2$
GeV and $\theta _{\pi ^0} \leq 16^0$ performed within the framework of
FRITIOF model. The calculation results are normalized on the
nucleus-nucleus interactions cross-sections obtained  in Glauber
approach  \cite{DIAGEN}.  As seen, the slopes of experimental and the
calculated curves are close, but the calculated cross-sections
overestimate the experimental values 2-3 times.

\begin{figure}[cbth]
\begin{center}
\psfig{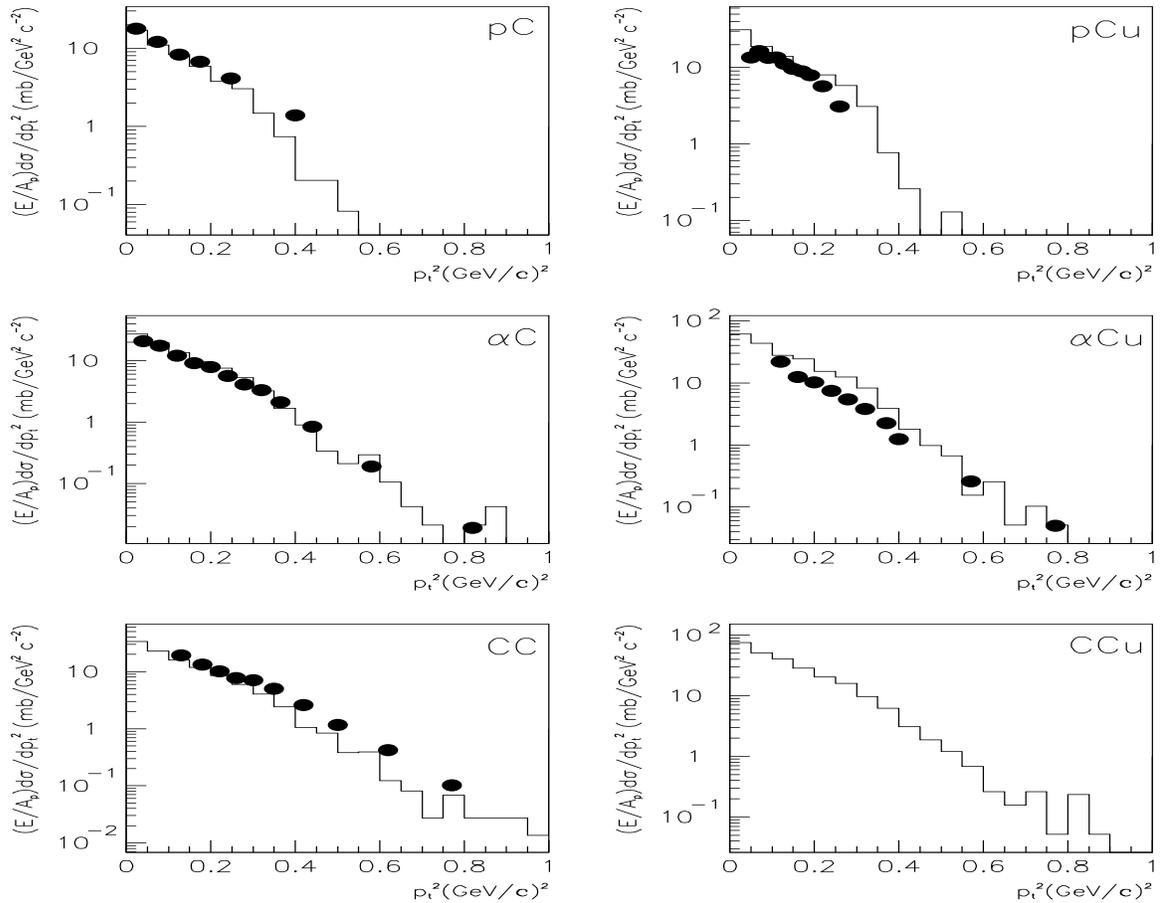}
\caption{Differential inclusive
cross-sections of $\pi ^0$-meson production in nucleus-nucleus
interactions
vs  transverse momenta
at 4.5 GeV/c/nucleon.  The notation is identical to that
in fig. 1.}
\end{center}
\label{fig2}
\end{figure}
Fig. 2 illustrates a better agreement between the calculations
and the experimental data. Fig. 2 gives the invariant
cross-sections of $\pi ^0$-mesons production  with respect to the $\pi
 ^0$ -meson transverse momentum.  The model reproduces both the spectrum
 forms and the absolute values of the cross-sections.  The reason of
 such different descriptions of the experimental data of fig. 1 and
fig.  2 is not clear for us.

The model FRITIOF allows one to decipher the cumulative
particle production mechanism in detail. The different characteristics
of $CC$-interaction events accompanied by  the
 fast $\pi ^0$ -meson production  are presented in fig. 3.  Fig.
 3a shows the yields into the invariant inclusive
 cross-section of projectile and target nucleons  ( dashed and dotted
 curves, respectively). The relative yields are given in fig. 3c.
   As
 seen, the contribution of the target nucleons is about $\sim$ 25 \%.
\begin{figure}[cbth]
\begin{center}
\psfig{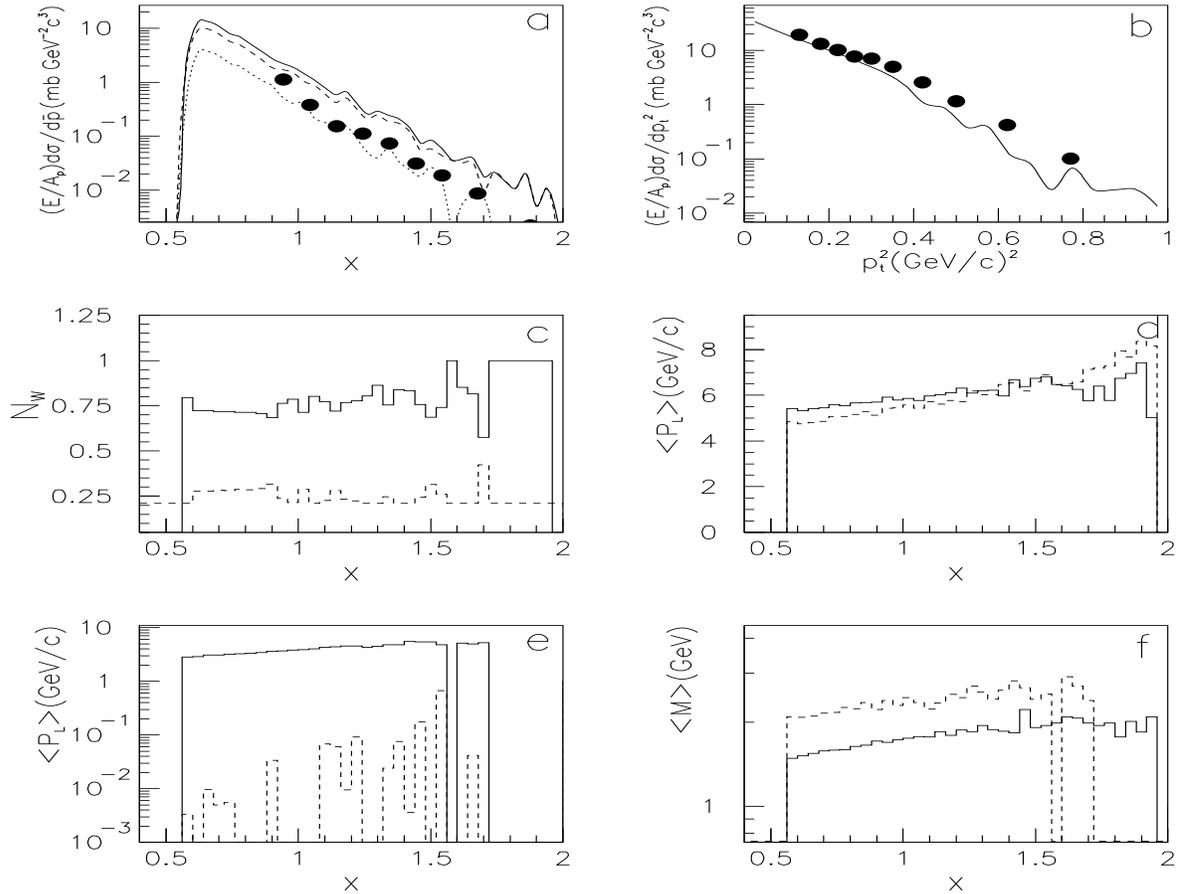}
\caption{Various characteristics of
$CC$-interactions at 4.5 GeV/c/nucleon with fast $\pi ^0$-meson
production.}
\end{center}
\label{fig3}
\end{figure}

Fig. 3d shows the average longitudinal momenta of projectile
nucleons before and after the interaction (solid and dashed curves,
respectively). According to the figure, more and more energetic
projectile nucleons are selected with increase the cumulative number.
Accounting the Fermi-motion is not critical for the description of
the inclusive cross-sections because without the Fermi-motion the
cross-section in the region of $X \sim$ 0.9 -- 1.3  does not decrease
in needed quantity, the slope of the cross-section is only changed (see
fig. 4). It is natural that the longitudinal momenta of
projectile nucleons decrease  during the interaction, but this
takes place below $X \sim 1.5$ (see fig. 3d). The nucleons acquired the
momenta larger than average momenta of incident nucleons, give the
contribution into the fast $\pi ^0$ -meson production in the region of
 large $X$.  It is clearly seen in the calculations performed without
taking the Fermi-motion into account (see insert in fig. 4, where the
solid line shows  the longitudinal momenta of projectile nucleons
before the interaction, dashed line shows the same after the
interaction with the fast $\pi ^0$ -meson production).  The considered
effect of the nucleon acceleration is a specific feature of the assumed
nucleus-nucleus interaction mechanism.

\begin{figure}[cbth]
\begin{center}
\psfig{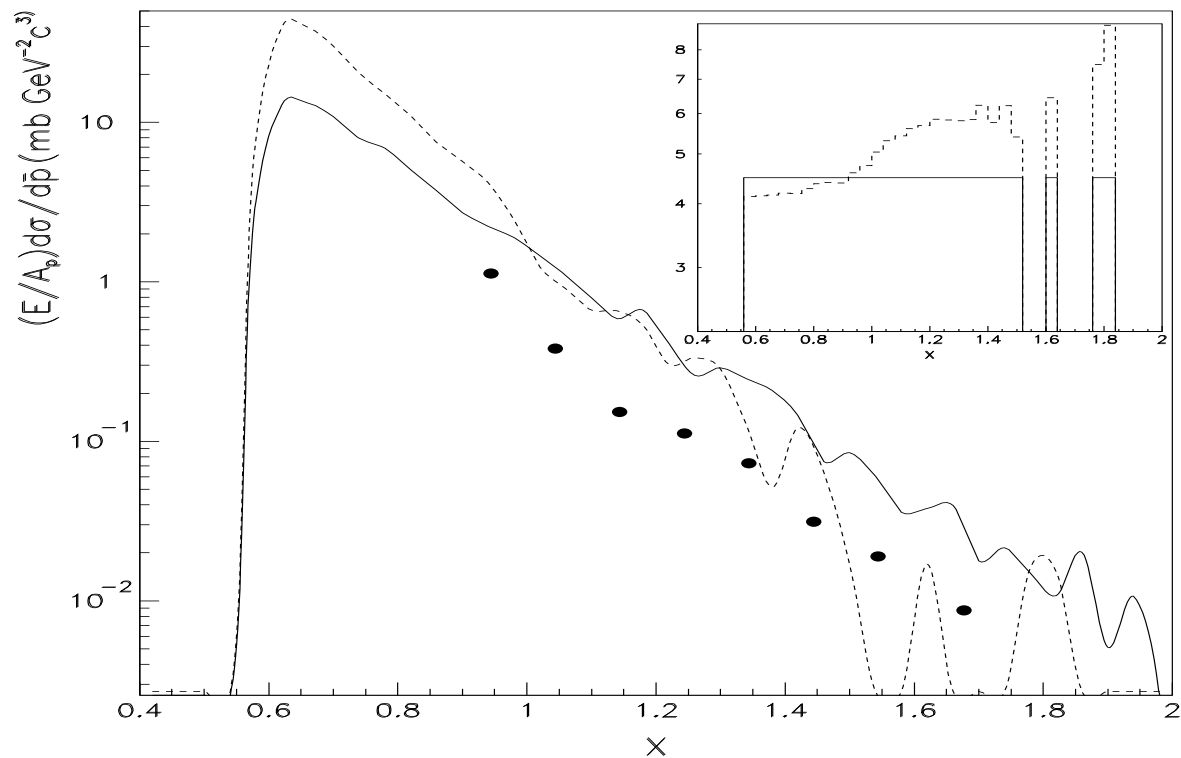}
\caption{Invariant inclusive
cross-sections of $\pi ^0$-meson production in $CC$ - interactions at
4.5 GeV/c/nucleon. The points are
 experimental data of the FOTON setup, the solid and dashed lines are
 the FRITIOF model calculations with and without fermi-motion,
 respectively.}
\end{center}
\label{fig4}
\end{figure}

Fig. 3e gives the analogous characteristics of the target
nucleons.  The longitudinal momenta of target nucleons are small before
the interaction (see dashed curve). In the course of the interaction,
the nucleons have to acquire significant longitudinal momenta for the
cumulative particle production in the forward direction (see solid
curve).

Fig. 3f shows the masses of the projectile and target nucleons
(see solid and dashed curves), which give the cumulative meson, after
the interaction.  As seen, the  target nucleons acquire larger
excitations than the  projectile nucleons. That coincide with the
main imaginations of the "hot" models with the exception of the
possibility of the nucleon acceleration.

The presented results allow one to expect a description of spectra of
  charged particles
  produced in nucleus-nucleus interactions
up to and out of kinematical
limit of free NN-collisions.

To test the possibility of applying model FRITIOF for reproduction of the
characteristics of cumulative charged particles,  we
consider the experimental data on $\pi ^-$ -meson production in
backward hemisphere \cite{PROPAN-PI}.  In the fig. 5 we
present the invariant inclusive cross-sections of  the $\pi ^-$ -meson
production in the $pC$-, $dC$-, $\alpha C$-, and $CC$-interactions at
momentum of projectile 4.2 GeV/c/nucleon  as a function of kinetic
energy.  There are  shown three groups of $\pi ^-$ -mesons: with
  emission angles  from 90$^\circ$ to 110$^\circ $, from 110$^\circ$ to 130$^\circ $,
from 130$^\circ$ to 180$^\circ $.
\begin{figure}[cbth]
 \begin{center}
 \psfig{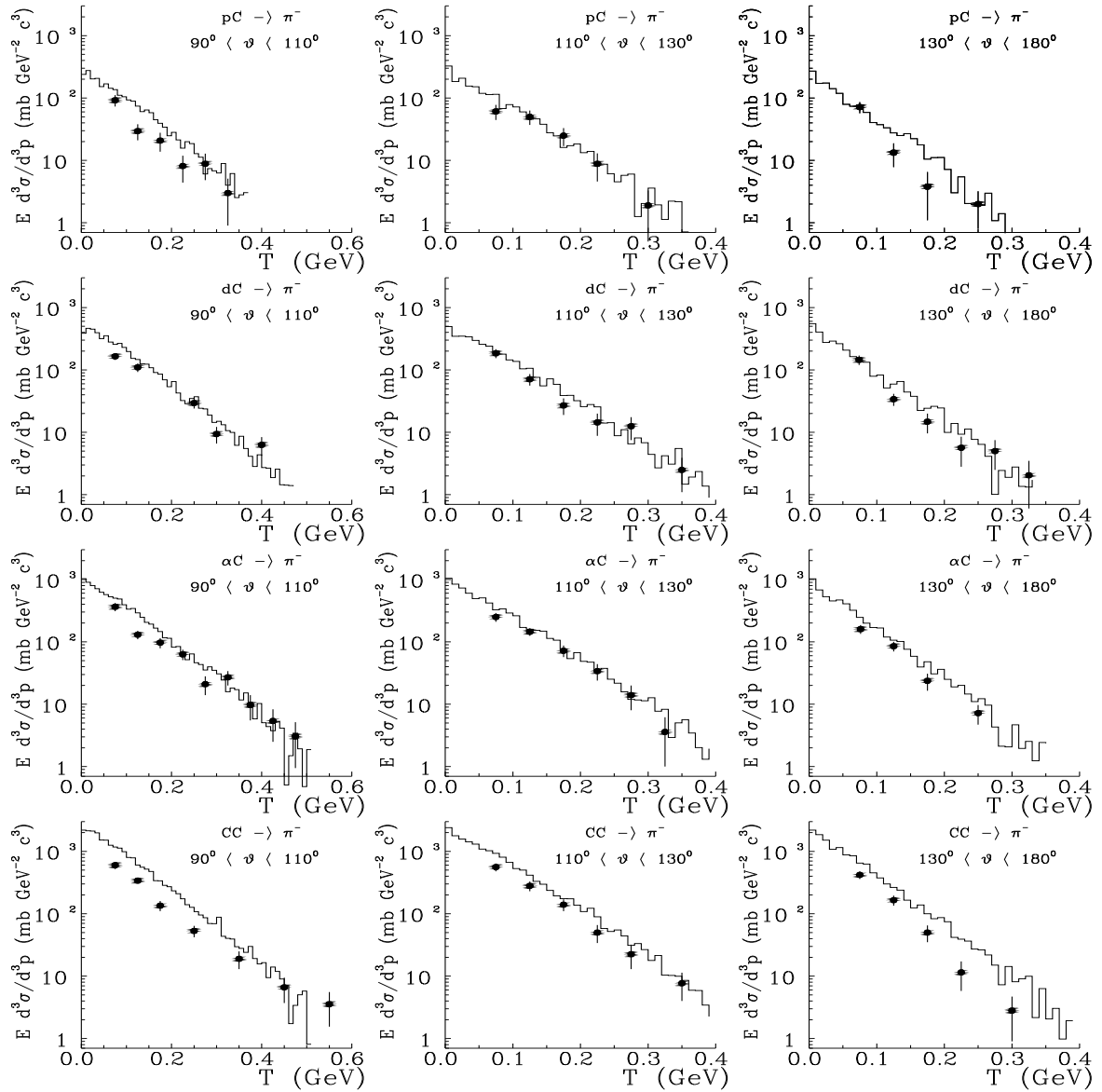}
 \caption{Invariant inclusive
  cross-sections of $\pi ^-$-meson production in $pC$-, $dC$-, $\alpha
  C$- and $CC$-interactions at 4.2 GeV/c/nucleon.  $T$- kinetic energy
  of $\pi ^-$-mesons, $\theta$ - angle of $\pi ^-$-meson production in
  laboratory system.
  The points are experimantal data of the propan collaboration,
  histograms are the FRITIOF model calculations.}
 \end{center}
\label{ch4_fig1}
\end{figure}

In the fig. 6, we present the
same for $dTa$-, $\alpha Ta$-, and $CTa$-collisions. As seen from the
figs. 5 and 6, the calculations in the frame of
 modified by V.Uzhinskii FRITIOF model \cite{UZHI-FRITI}
  reproduce qualitatively (in
the some case quantitatively) the dependence of the invariant
cross-section of the  $\pi ^-$- meson production on kinetic energy, on
the mass of projectile nucleus, on the mass of target nucleus, and on
the emission angle.
\begin{figure}[cbth]
\begin{center}
\psfig{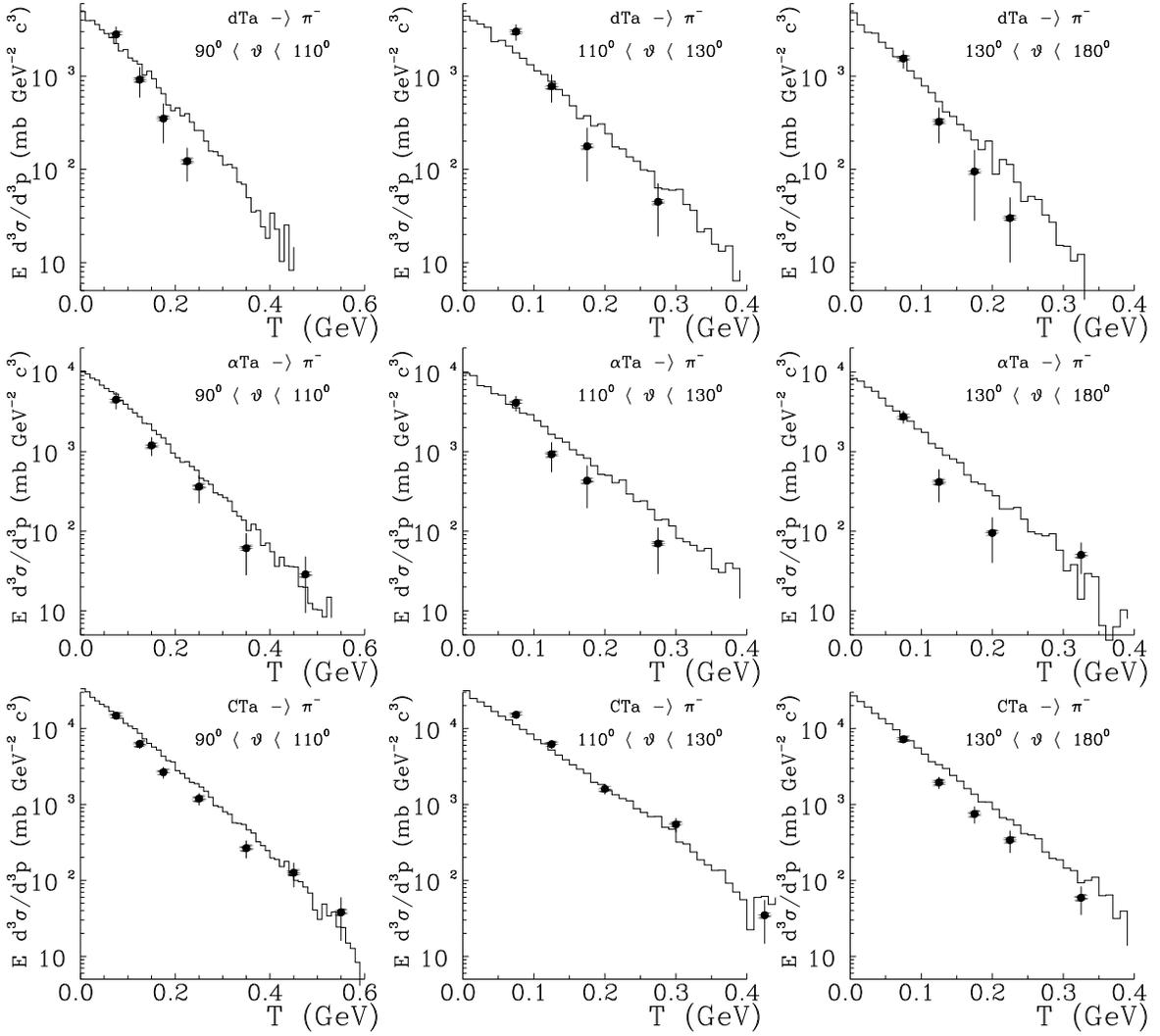}
\caption{Invariant inclusive
cross-sections of $\pi ^-$-meson production in  $dTa$-, $\alpha
Ta$- and $CTa$-interactions at 4.2 GeV/c/nucleon.  $T$- kinetical energy
of $\pi ^-$-mesons, $\theta$ - angle of $\pi ^-$-meson production in
laboratory system.  The points are experimantal data
of the propan collaboration, histograms are the FRITIOF model calculations.}
\end{center}
\label{ch4_fig2}
\end{figure}

For  more detail consideration, let us turn to spectra of
 the  $\pi ^-$ -meson production in the $NN$-, $pC$-, and
$CC$-interactions in the backward hemisphere \cite{PROPAN-PI}
(fig. 7).
According to the calculations, spectra of $\pi ^-$ -mesons in the  $NN$
and nucleus-nucleus interactions are similar.  The description of pion
spectra in the $AA$-interactions was achieved without free, fitting
parameters. In  given approach, the similarity of pion spectra is
explained easily.  In the nucleon-nucleon collisions at sufficiently
high energies, spectra in the region of target fragmentation are not
dependent on the interaction energy. In the hadron-nucleus
interactions, participating nucleons fragment independently
on projectile particle, as in $NN$-collisions.
Therefore, every participant-nucleon gives independent contribution
to the cross-section. Mean multiplicity of such nucleons is proportional
$<\nu > \sim A^{1/3}$. Then, the process cross-section
is proportional
$\sigma \sim <\nu > \sigma ^{in}_{hA} \sim A^{1/3} A^{2/3} = A^1.$

In the nucleus-nucleus interactions, the similarity of spectra
can take place, if target nucleons collide no more than one time.
The calculations show that in the considered interactions of
light nucleus with light nucleus and heavy nucleus, it is really so.
\vspace{-0.3cm}
\begin{figure}[cbth]
\begin{center}
\psfig{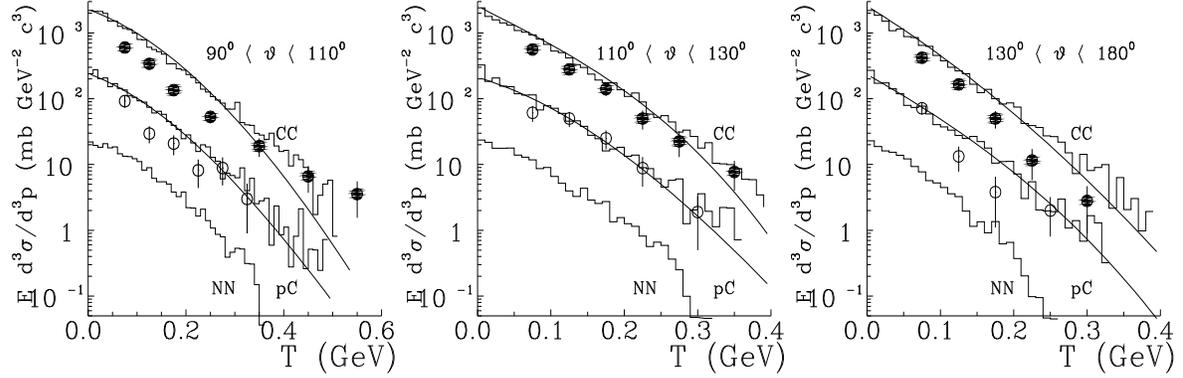}
\caption{Invariant inclusive cross-sections of $\pi ^-$-meson production
in  $NN$-, $pC$- ¨ $CC$-interactions.
The points are experimental data of the propan collaboration,
histograms are the FRITIOF model calculations,
curves are calculations without fermi-motion.}
\end{center}
\label{ch4_fig3}
\end{figure}
The spectra of mesons can be considered as similar only in the first
rough approximation. More detailed consideration shows that it isn't
possible  to describe meson spectra in NN-collisions by simple
exponent. At the same time, spectra of AA-interactions can be described
by exponent quite well.  The difference is caused by the interaction
mechanism.  It is obviously, the main distinction is so that spectra of
$NN$-collisions are restricted by allowed kinematical region, and
spectra of $AA$-interactions extend over this region. For presented
experimental data, only last points are out of kinematical limit of
free $NN$-interactions. According to the fig. 7
spectra of $\pi ^-$-mesons change unessentially with taking into
account fermi-motion of nucleus in kinematical region
allowed in NN-interactions.
Production of mesons without fermi-motion of nucleos  out of
this region is described by mechanism of the nucleon mass increase.

\begin{figure}[cbth]
\begin{center}
\psfig{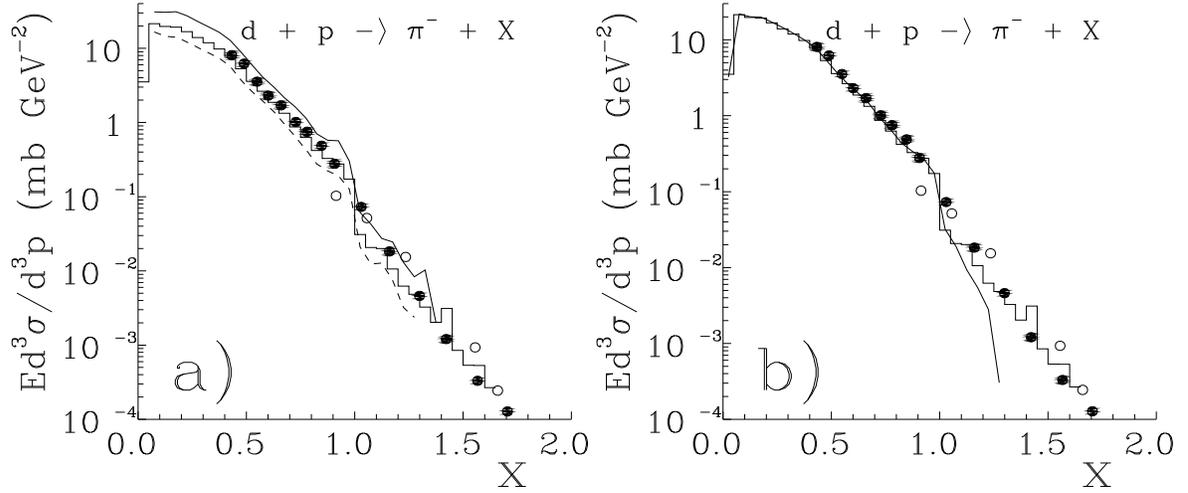}
\caption{Invariant inclusive cross-sections of $\pi ^-$-meson production
in   $dp$-interactions.
The points are experimental data of the SPHERA collaboration,
curves are the FRITIOF model calculations.}
\end{center}
\label{ch4_fig5}
\end{figure}

Let us consider dependence of the cumulative particle production
cross-section on target mass number. For this let us turn to the data
of $dA$-reaction \cite{litv}  at momentum of
projectile deuteron equals to
8,9~GeV/c with production of $\pi ^-$-mesons at zero degree.
The fig. 8 shows the cross-section as a function
of variable $X$. Variable $X$  is determined as
$$
x=\frac {M_N E_{\pi ^-}-\frac{1}{2} M_{\pi ^-}^2}
{E_N M_N-E_NE_{\pi ^-}- M_N^2+P_NP_{\pi ^-} \cos \theta _{\pi ^-}},
$$
where $M_N$, $M_{\pi ^-}$, $P_N$, $P_{\pi ^-}$, $E_N$, $E_{\pi ^-}$ are
masses, momenta and energies of nucleon and $\pi ^-$-meson respectively;
$\theta _{\pi ^-}$ -- emission angle of $\pi ^-$-mesons in lab.
system.
In the fig. 8, the solid lines is the standard model
calculation. The shape of the calculated curve is near to the
experimental data, but the calculations overestimate the experimental
data.  The simplest way to reach an agreement between the experiment
and the theory is  a decrease of NN-interaction cross-section
in nucleus.  At the
 50$\%$ decrease the calculations underestimate the experimental data
 (see the dashed lines), and at the 30$\% $ decrease (see the
 histograms in the fig. 8) we have a good agreement. The other
 possibility to reach the agreement is a variation of  model
 parameters, that leads to the change of the cross-section shape.  For
 example, the solid lines in the fig. 8b give the calculations without
 de-excitation of the nucleons.  At the 30$\%$ decrease of
 $NN$-interaction cross-section in the nucleus, we have also a good
 description of the cross-section dependence on target mass number.

  The fig. 9 gives a ratio of  deuteron-nucleus and deuteron-proton
 cross section. The points are experimental data at $X$ equal of 1.23
 \cite{litv}, the histograms are calculations.
\begin{figure}[cbth]
 \begin{center}
\psfig{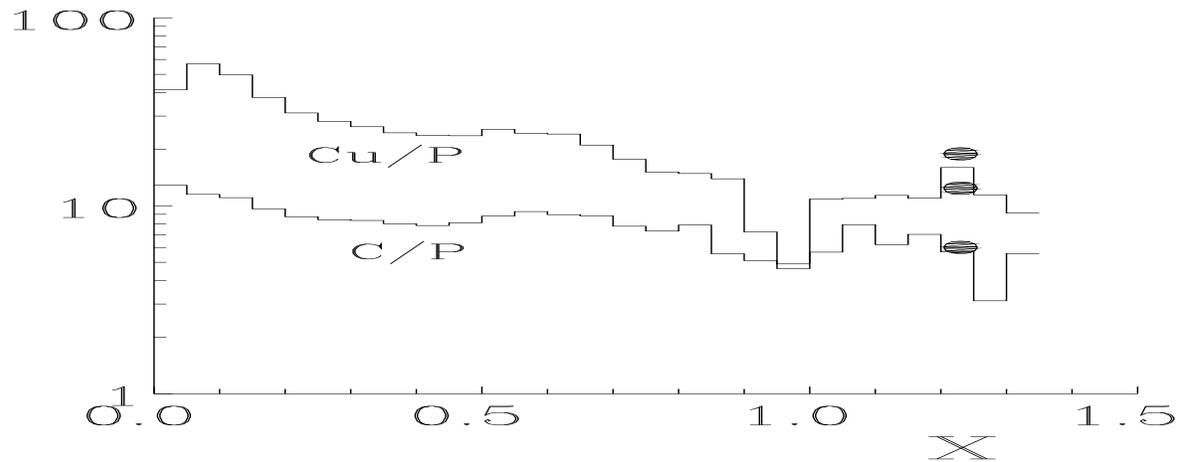}
\caption{Ratio of cross-sections of $\pi ^-$-meson production on
the nucleus and on the proton  at x=1.23.
The points are experimental data of the SPHERA collaboration,
curves are the FRITIOF model calculations.}
\end{center}
\label{ch4_fig6}
\end{figure}
 The weak A-dependence is explained
 by the following effect.
 Our calculations show that the $dA$-collisions  with two NN-interactions
 give the main contributions to the cumulative particle spectra.
 In the collisions with the large number of the NN-interactions
 there is not practically cumulative particles due to energy losses of
 projectile deuterons. In the collisions with two NN-interactions, the
 processes dominate where two projectile nucleons collide with one
 target nucleon. These processes have peripheral character.

 The more complete situation takes place with a description of
 cumulative proton spectra.
\begin{figure}[cbth]
\begin{center}
\psfig{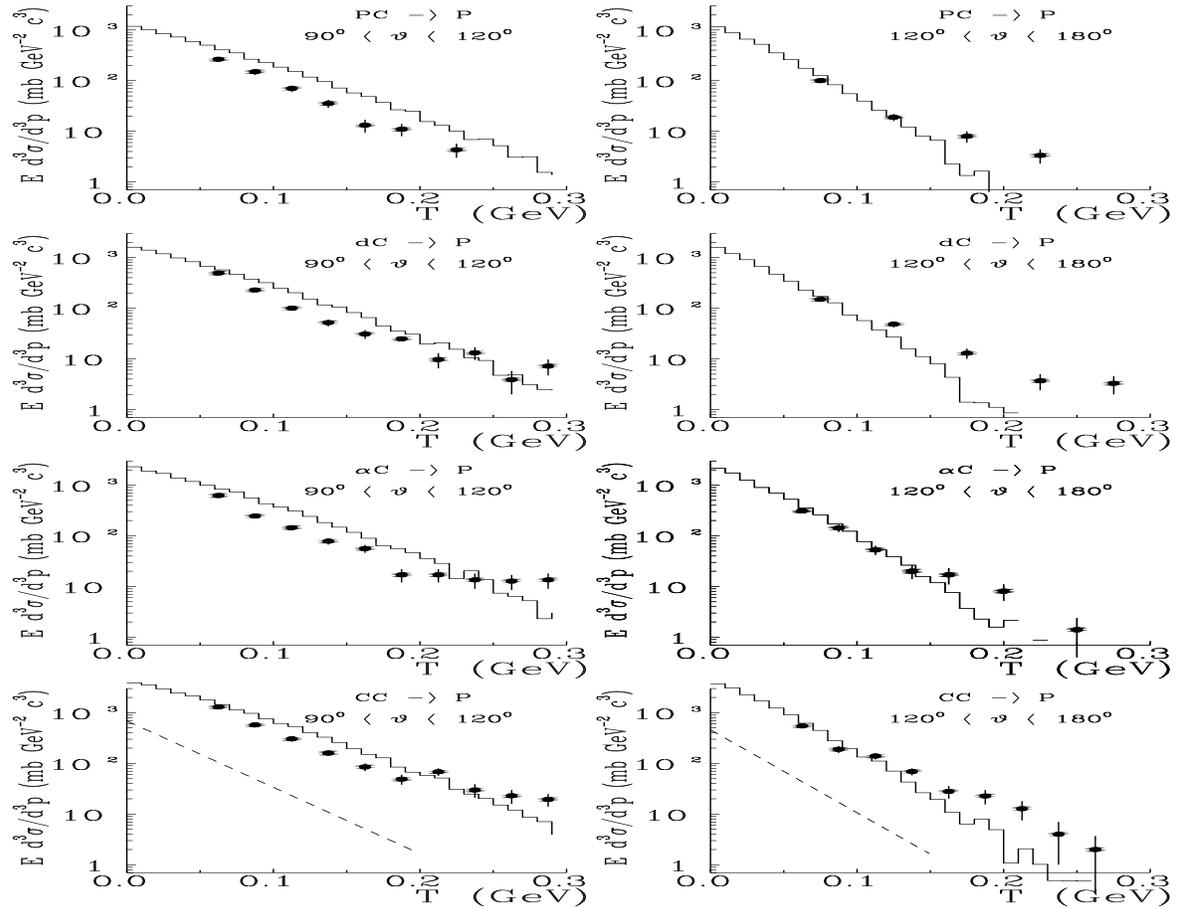}
\caption{Invariant inclusive cross-sections of proton production
in  $AC$-interactions.
The points are experimental data of the propan collaboration,
histograms are the FRITIOF model calculations.}
\end{center}
\label{ch4_fig8}
\end{figure}
  In  fig. 10 the invariant
 inclusive cross-sections of protons emitted in backward hemisphere are
 presented as a function of kinetic energy. Points are
 propan collaboration experimental data \cite{PROPAN-P}, histograms are
 our calculations. The shape of the curves is determined by
 Fermi-motion, and the absolute values of the cross-sections is
 determined by nuclear destructure model. The reggeon theory inspired
 model of nuclear destructure is used in modified
 FRITIOF model \cite{GAL-UZHI}.  It is assumed, that each
 intra-nuclear collision initiate reggeon exchanges in the spectator
 part of the nucleus. The probability to evolve spectator nucleon to
 the cascade is given as
 $$\protect \Large W(\vec{b})=C_{nd}
  exp(\vec{b} ^2/r^2_{nd}), $$
   where $b_{ij}$ is a difference between
impact coordinates of the spectator nucleon and inelastic interacted
nucleon. $C_{nd}$ and $r_{nd}$ are parameters.  Fitting these
parameters, we have a good description only for the "soft" part of the
proton spectra. For $AC$-interactions, we choose $C_{nd}=1$, and
$r_{nd}^2=1.1~(fm^2)$. But
for $ATa$-interactions \cite{CTa}, we have $C_{nd}=0.2$ and $r_{nd}
^2=1.1~(fm^2)$ (see fig. 11).
\begin{figure}[cbth]
\begin{center}
\psfig{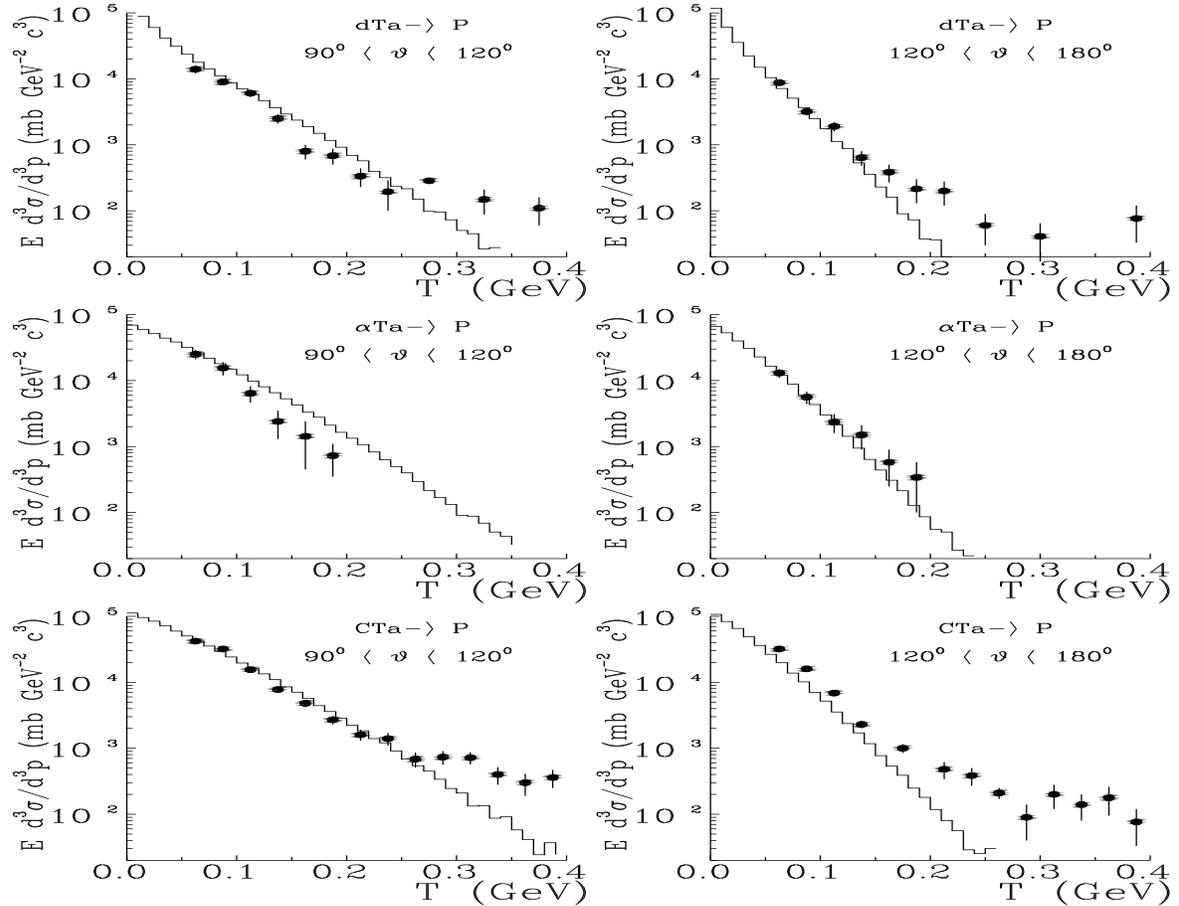}
\caption{Invariant inclusive cross-sections of proton production
in  $ATa$-interactions.
The points are experimental data of the propan collaboration,
histograms are the FRITIOF model calculations.}
\end{center}
\label{ch4_fig11}
\end{figure}

The experimental data \cite{PROPAN-P},\cite{CTa} show a change of a
slope of proton spectra at $T \sim 0.2-0.25$~GeV.
 There are not such changes in the model calculations.
 At the same time, we reproduce the dependence of the
cross-sections on the emission angle of protons, and projectile and
target mass numbers. For explanation of hard part of the cumulative
proton spectra, it is needed to use the other production mechanism (for
example multi-quark bags, fluctons, and so ones).

The author express her sincere gratitude to  V.V. Uzhinskii,
G.L. Melkumov, E.N. Kladnickaya
and A.G. Litvinenko for their fruitful discussions and
valuable remarks.

\end{document}